\documentclass{aipproc}

\layoutstyle{6x9}

\SetInternalRegister\hbadness{8000} 

\newcommand\doingARLO[2][]{%
  \ifx\mmref\undefined #1\else #2\fi
}
\begin{document}

\begin{flushright}
Preprint DFPD 01/TH/17\\
hep-th/0105102
\end{flushright}

\title{Introduction to the Superembedding Description of Superbranes }

\classification{PACS: 11.15. - q; 11.17. + y}
\keywords{Superparticles, superstrings, supergravity, branes, surface theory}

\author{Dmitri Sorokin}{
  address={INFN, Sezione di Padova, via F. Marzolo 8, 35131, Padova, Italia\\
  and Institute for Theoretical Physics, National Science Center, KIPT,
Kharkov, Ukraine},
  email={sorokin@pd.infn.it},
 thanks={
}
}

\copyrightyear{2001}

\begin{abstract}
Basics of the geometrical formulation of the dynamics of
supersymmetric objects are considered and its relation to
conventional formulations of superbranes is discussed. In
particular, we demonstrate how the kappa--symmetry of the
Green--Schwarz formulation shows up from local worldvolume
supersymmetry, and briefly discuss applications of the
superembedding approach.
\end{abstract}

\date{\today}

\maketitle

\section{Introduction}
Superembedding is an elegant and geometrically profound approach
which is based on a supersymmetric extension of the classical
surface theory to the description of superbrane dynamics by means
of embedding worldvolume supersurfaces into target superspaces
(see \cite{soro-1} for a detailed review).

The superembedding approach  arose as a proposal to solve the
problem of the covariant quantization of the Green--Schwarz
superstring by combining in a more general formulation main
properties of both the Neveu--Schwarz--Ramond (NSR) and the
Green--Schwarz (GS) formulation, and by now it has developed into
a generic geometrical method for formulating the theory of
superbranes.

Being manifestly doubly supersymmetric (on the worldvolume and in
target superspace) the superembedding approach has explained the
origin and the nature of the fermionic $\kappa$--symmetry of the
GS formulation as a manifestation of the conventional local
supersymmetry of the worldvolume, and thus solved the problem of
infinite reducibility of the $\kappa$--symmetry by realizing it as
an irreducible extended worldvolume supersymmetry \cite{soro-stv}.
This stimulated progress in the covariant quantization of the
Green--Schwarz superstring mainly due to persistent work of
N.~Berkovits \cite{soro-berquant}.

The superembedding approach established or clarified a classical
relationship between various formulations of the dynamics of
superparticles and superstrings, such as the NSR and the GS
formulation \cite{soro-nsr}, the twistor \cite{soro-twist} and
harmonic \cite{soro-harm} descriptions.

This approach has proved to be a universal and powerful method
applicable to the description of all known supersymmetric branes,
in particular, to those of them for which standard methods
encountered problems because of their specific structure, such as
the 5--brane of M--theory. The superembedding methods allowed, for
the first time, to derive the complete set of covariant equations
of motion of the M5--brane \cite{soro-hsw}. And only later these
equations were obtained \cite{soro-M5e} from the M5--brane action
\cite{soro-M5a} based on a different technique adapted to deal with
self--dual fields.

The superembedding formulation has also proved to be useful for
studying the ``brany'' mechanism of partial supersymmetry breaking
\cite{soro-brany} by giving a geometrical recipe \cite{soro-pst,soro-bppst,soro-dh}
for constructing covariant worldvolume supersymmetric actions for
superbranes. Upon gauge fixing worldvolume superdiffeomorphisms
these actions become those of effective field theories with
non--linearly realized spontaneously broken supersymmetries. This
has demonstrated an intrinsic link of the superembedding approach
and the method of nonlinear realizations developed in application
to supersymmetric theories in \cite{soro-pbgs} (see
\cite{soro-bik} for a review and references).

In this contribution I would like to make an introduction into the
superembedding formalism and describe its basic properties.

\section{Double Supersymmetry}
As it has been mentioned in the Introduction the superembedding
description of superbrane dynamics is formulated in such a way
that it possesses manifest supersymmetry on the brane worldvolume
and in target superspace, and thus has properties of both the NSR
and the GS formulation.

\subsection{The Neveu--Schwarz--Ramond formulation}
In the NSR formulation, which describes spinning particles
\cite{soro-sp} and spinning strings~\cite{soro-ss}\footnote{No spinning branes
have been consistently constructed so far \cite{soro-ht}.}, the
worldline or worldsheet of the spinning object is a supersurface
${\cal M}_w$ parametrized by bosonic coordinates $\xi$ and
fermionic coordinates $\eta$ which we will collectively call
$z^M=(\xi,\eta)$.  Depending on the model considered ${\cal M}_w$ may have
a various number of fermionic directions $\eta$. For simplicity we
here take only one $\eta$. The dynamics of the spinning object is
described by embedding ${\cal M}_w$ into a bosonic target
space--time ${M_T}$ parametrized by coordinates $X^{\underline m}$
($\underline m =0,1,\cdots,D-1$). In the classical problems the
number of space--time dimensions can be arbitrary, but for quantum
consistency the spinning string, whose worldsheet has one or two
fermionic directions, must live in a ten--dimensional target
space.

The motion of the spinning object is described by the image of
${\cal M}_w$ in $M_T$
\begin{equation}\label{soro-1}
X^{\underline m}(z^M)=x^{\underline m}(\xi)+i\eta\chi^{\underline
m}(\xi),
\end{equation}
where $x^{\underline m}(\xi)$ is associated with the bosonic
degrees of freedom of the spinning particle or string in ${M_T}$,
and the Grassmann--odd vector
$\chi^{\underline m}(\xi)$ is associated with its spin degrees of freedom.

The NSR formulation is invariant under worldsheet
superdiffeomorphisms $z^M~\rightarrow~z^{'M}(z^M)$, which include
${\cal M}_w$ bosonic reparametrizations
\begin{equation}\label{soro-2}
\delta\xi=a(\xi),
\end{equation}
and local worldsheet supersymmetry
\begin{equation}\label{soro-3}
\delta\eta=\kappa(\xi), \qquad \delta\xi=i\kappa(\xi)\eta.
\end{equation}

The presence of the local symmetries (\ref{soro-2}) and
(\ref{soro-3}) implies that the dynamics of the spinning objects
is subject to bosonic and fermionic first--class constraints,
respectively, which I will {\it schematically} write down in the
form
\begin{equation}\label{soro-4}
(\partial x^{\underline m})^2=0, \qquad \partial x^{\underline
m}\chi_{\underline m}=0.
\end{equation}
The bosonic constraint in (\ref{soro-4}) is the mass shell or the
Virasoro conditions, and the fermionic constraint produces, upon
quantization, the Dirac equation for the spin wave functions of
the dynamical system.

The NSR formulation does not have target--space supersymmetry. The
latter appears, in the case of the spinning string, only at the
quantum level upon imposing the Gliozzi--Scherk--Olive projection.
The merit of this formulation is that it is covariantly
quantizable.

\subsection{The Green--Schwarz formulation}
This formulation is applicable to all known superparticles,
superstrings  and superbranes\footnote{The name ``Green--Schwarz"
used for this formulation should be regarded as cumulative, since
for different extended objects it has been developed by different
people. This also concerns the ``NSR'' fromulation. A detailed
list of references the reader may find in
\cite{soro-1}.}.

Now the worldvolume ${\cal M}_w$ is a (p+1)--dimesnional bosonic
surface parametrized by the coordinates $\xi^m$ ($m=0,1,\cdots,p$)
and the target space, into which ${\cal M}_w$ is embedded, is a
superspace
$M_{TS}$ parametrized by bosonic coordinates $x^{\underline m}$
($\underline m =0,1,\cdots,D-1$) and by an appropriate number of
fermionic coordinates
$\theta^{\underline\alpha}$ ($\underline\alpha =1,\cdots,2n$)
\begin{equation}\label{soro-5}
Z_{TS}^{\underline M}(\xi)=\left(x^{\underline
m}(\xi),~~~\theta^{\underline\alpha}(\xi)\right).
\end{equation}
The GS formulation is manifestly invariant under bosonic ${\cal
M}_w$ reparametrizations $\xi~ \rightarrow ~\xi'(\xi)$, and target
space superdiffeomorphisms $Z^{\underline
M}~\rightarrow~Z^{'\underline M}(Z^{\underline M})$, which in the
case of flat target superspace reduce to the translations along
$x^{\underline m}$ and to the global target--space supersymmetry
transformations
\begin{equation}\label{soro-6}
\delta\theta^{\underline\alpha}=\epsilon^{\underline\alpha},
\qquad
\delta x^{\underline
m}=i\bar\theta\Gamma^{\underline m}\delta\theta.
\end{equation}
There is also another (non--manifest) local worldvolume fermionic
symmetry, so--called $\kappa$--symmetry, inherent to the GS
formulation. This is an important symmetry which implies and
reflects the existence of supersymmetric BPS brane--like solutions
of corresponding supergravity theories. It is thus responsible for
the `brane scan' (i.e. which brane lives in which target
superspace). The $\kappa$--symmetry was first observed in the case
of superparticles
\cite{soro-k} and has the following generic form of transformations (in
flat target superspace):
\begin{equation}\label{soro-7}
\delta\theta^{\underline\alpha}=\Pi^{\underline\alpha}_{~~\underline\beta}
\kappa^{\underline\beta}(\xi),
\qquad
\delta x^{\underline
m}=-i\bar\theta\Gamma^{\underline m}\delta\theta,
\end{equation}
where $\Pi^{\underline\alpha}_{~~\underline\beta}$ is a projector
matrix ($\det{\Pi^{\underline\alpha}_{~~\underline\beta}}=0$),
whose form depends on the object under consideration.

Because of the presence of the projector in the
$\kappa$--transformations they are infinite reducible, so
only half of $\kappa^{\underline\alpha}$, i.e. $n$ of the
$2n$ Grassmann spinor components, effectively contribute to
the variation of the worldvolume fields. This is a cause of the
problem of the covariant quantization of the GS formulation.

To solve the infinite reducibility problem of the
$\kappa$--symmetry it is natural to try to find its irreducible
realization which is covariant in target superspace. A natural
assumption is that this should be an extended local worldvolume
supersymmetry (\ref{soro-3}) with the number of independent
parameters equal to the number of independent (irreducible)
$\kappa$--symmetries \cite{soro-stv}. This reasoning brings us to

\subsection{The doubly--supersymmetric formulation}

In this formulation the dynamics of superbranes is described by
embedding a worldvolume {\sl super}surface ${\cal M}_{sw}$
parametrized by the coordinates $z^M=(\xi^m,\eta^\alpha)$
($m=0,1,\cdots,p$), ($\alpha=1,\cdots, n)$ into a target {\sl
super}space
$M_{TS}$ parametrized by the coordinates $Z^{\underline M}=(X^{\underline m},
\Theta^{\underline\alpha})$ ($\underline m=0,1,\cdots,D-1$), ($\underline\alpha=1,...,2n$).
Note that the number of the Grassmann
directions of ${\cal M}_{sw}$ is half the number of the Grassmann
directions of $M_{TS}$. Such a choice of the supermanifolds for
superembedding is caused by our desire to identify $n$ local
supersymmetries on ${\cal M}_{sw}$ with $n$ independent
$\kappa$--symmetries of the GS formulation.

Thus in the doubly supersymmetric formulation the degrees of
freedom of the superbranes are described by worldvolume
superfields
\begin{equation}\label{soro-8}
X^{\underline m}(z^M)=x^{\underline
m}(\xi)+i\eta^\alpha\chi^{\underline m}_\alpha(\xi)+~\cdots, \quad
\Theta^{\underline\alpha}(z^M)=\theta^{\underline\alpha}(\xi)
+\eta^\alpha\lambda^{\underline\alpha}_\alpha(\xi)+~\cdots,
\end{equation}
where $\cdots$ stand for the terms of higher order in
$\eta^\alpha$. These terms contain auxiliary fields and in addition,
for example in the case of the D--branes and the M5--brane,
include the gauge fields propagating on the worldvolumes of these
branes.

From eq. (\ref{soro-8}) we see that in the doubly supersymmetric
construction the number of degrees of freedom of the superbrane
roughly speaking doubles. We now have $x^{\underline m}(\xi)$
describing the bosonic oscillations of the brane, the Grassmann
`spin'--vectors $\chi^{\underline m}_\alpha(\xi)$ as in the NSR
formulation, the Green--Schwarz fermionic spinor degrees of
freedom $\theta^{\underline\alpha}(\xi)$ and their bosonic
counterparts $\lambda^{\underline\alpha}_\alpha(\xi)$. So if all
these worldvolume fields are independent the corresponding models
will not describe conventional superbranes. In this case one will
get, for instance, so called spinning superparticles and spinning
superstrings \cite{soro-spins} which have more degrees of freedom
then the conventional NSR and GS dynamical systems. They also have
infinite reducible $\kappa$--symmetry as a fermionic symmetry
independent of the local worldvolume supersymmetry.

To reach our goal of interpreting $\kappa$--symmetry as a
manifestation of the local worldvolume supersymmetry we should
find an appropriate doubly--supersymmetric description of the
{\it conventional} superbranes. For this we should impose constraints on the
superfields (\ref{soro-8}) which relate their components in such a
way that the independent physical degrees of freedom described by
these superfields will correspond to the standard GS formulation.
The geometrical meaning of these constraints is that they cause
the worldvolume supersurface ${\cal M}_{sw}$ to be imbedded into
the target superspace ${M}_{TS}$ in a specific way.

\section{The Superembedding Condition}
Before discussing in more detail the geometrical meaning of the
superembedding condition let us consider its dynamical
consequences in the simplest case of a superparticle propagating
in a flat $N=1$, $D=3$ target superspace. Then the supersurface
${\cal M}_{sw}$ of the previous subsection is associated with the
superparticle ``worldline" having one bosonic (time) and one
fermionic coordinate $(\xi,\eta)$, and the target superspace is
parametrized by bosonic three--vector coordinates $X^{\underline
m}$ ($m=0,1,2$) and Grassmann Majorana two--spinor coordinates
$\Theta^{\underline\alpha}$ $(\underline\alpha=1,2)$. The
superembedding condition relates the worldline superfields
$X^{\underline m}(z^M)$ and $\Theta^{\underline\alpha}(z^M)$ in
the following way
\begin{equation}\label{soro-9}
DX^{\underline m}-iD\bar\Theta\Gamma^{\underline m}\Theta=0,
\end{equation}
where $D$ is a Grassmann covariant derivative on ${\cal M}_{sw}$
which in the case of the superparticles can be chosen to be flat
\begin{equation}\label{soro-10}
D={\partial\over{\partial\eta}}+i\eta{\partial\over{\partial\xi}},
\qquad \{D,D\}=2i\partial_\xi.
\end{equation}
Using the $\eta$--expansion (\ref{soro-8}) which in the case under
consideration does not contain the ``$\cdots$"--terms we obtain
the following relation between the components of the superfields
$X^{\underline m}(z^M)$ and $\Theta^{\underline\alpha}(z^M)$
\begin{equation}\label{soro-11}
\partial_\xi x^{\underline
m}-i\partial_\xi\bar\theta\Gamma^{\underline
m}\theta=\bar\lambda\Gamma^{\underline m}\lambda,
\end{equation}
\begin{equation}\label{soro-12}
\chi^{\underline m}=\bar\theta\Gamma^{\underline
m}\lambda.
\end{equation}
From eq. (\ref{soro-11}) we see that
$\lambda^{\underline\alpha}(\xi)$ are not independent fields and
are expressed in terms of the derivatives of $x^{\underline m}$
and $\theta^{\underline\alpha}$. Moreover in the l.h.s. of
(\ref{soro-11}) one can recognize the canonical momentum of the
superparticle $\Pi^{\underline m}=\partial_\xi x^{\underline
m}-i\partial_\xi\bar\theta\Gamma^{\underline m}\theta$ whose
square is identically zero
\begin{equation}\label{soro-13}
\Pi^{\underline m}\Pi_{\underline m}=0,
\end{equation}
 because of its so called Cartan--Penrose (or twistor)
representation as a bilinear combination of commuting spinor
components and $\Gamma$--matrix identities. We conclude that the
superparticle is massless.

{\sl In the case of the superstrings the superembedding condition will
produce in a similar way the Virasoro constraints, and it will
produce the corresponding constraints for the other superbranes.}

Eq. (\ref{soro-12}) implies the relation between the Grassmann
vector and the Grassmann spinor variables, so that only one or
another can be regarded as describing independent fermionic
degrees of freedom. This is a basic relation which allows one to
establish a classical correspondence between the NSR and the GS
formulation of supersymmetric particles and strings
\cite{soro-nsr}.

\subsubsection{Local worldvolume supersymmetry versus
$\kappa$--symmetry}

Let us now demonstrate how $\kappa$--symmetry appears in the
superembedding formulation as a weird realization of the local
worldvolume supersymmetry.

The components (\ref{soro-7}) of the superfields $X^{\underline
m}(z^M)$ and
$\Theta^{\underline\alpha}(z^M)$ transform under the local
worldline supersymmetry (\ref{soro-3}) in the standard way
\begin{equation}\label{soro-14}
\delta\theta^{\underline\alpha}=-\lambda^{\underline\alpha}\kappa(\xi),
\qquad\delta
\lambda^{\underline\alpha}=i\partial_\xi\theta^{\underline\alpha}\kappa(\xi),
\end{equation}
\begin{equation}\label{soro-15}
\delta x^{\underline m}=i\chi^{\underline m}\kappa(\xi), \quad
\qquad \delta\chi^{\underline m}=-\partial_\xi x^{\underline
m}\kappa (\xi).
\end{equation}
We now substitute into the first equation of (\ref{soro-15}) the
solution (\ref{soro-12}) of the superembedding condition
(\ref{soro-9}) and observe that, due to the form of the
$\theta$--variation (\ref{soro-14}), the variation of
$x^{\underline m}$ can be rewritten as follows
\begin{equation}\label{soro-16}
\delta x^{\underline m}=i\bar\theta\Gamma^{\underline
m}\lambda\kappa(\xi)=-i\bar\theta\Gamma^{\underline
m}\delta\theta.
\end{equation}

The next step is to replace the Grassmann scalar parameter of the
local supersymmetry by the scalar product of the spinor
$\lambda_{\underline\beta}$ with a Grassmann spinor parameter
$\kappa^{\underline\beta}(\xi)$, which is always possible,
\begin{equation}\label{soro-17}
\kappa(\xi)=2\lambda_{\underline\beta}\kappa^{\underline\beta}(\xi),
\end{equation}
and to substitute (\ref{soro-17}) into the $\theta$--variation
(\ref{soro-14}). We thus get
\begin{equation}\label{soro-18}
\delta\theta^{\underline\alpha}=
-2\lambda^{\underline\alpha}\lambda_{\underline\beta}\kappa^{\underline\beta}(\xi).
\end{equation}
We now note that, due to the superembedding condition
(\ref{soro-11}), the bilinear combination of
$\lambda$ in (\ref{soro-18}) is nothing but
\begin{equation}\label{soro-19}
\Pi^{\underline{\alpha}}_{~~\underline\beta}=
-2\lambda^{\underline\alpha}\lambda_{\underline\beta}=(\partial_\xi
x^{\underline m}-i\partial_\xi\bar\theta\Gamma^{\underline
m}\theta)(\Gamma_{\underline
m})^{\underline\alpha}_{~~\underline\beta},
\end{equation}
which is the projector matrix in the $\kappa$--symmetry variation
of $\theta$ (\ref{soro-7}). Hence, the local supersymmetry
variations (\ref{soro-16}) and (\ref{soro-18}) reduce to the
$\kappa$--variations (\ref{soro-7}).

{\sl We have thus demonstrated how, in virtue of the superembedding
condition (\ref{soro-9}), the
$\kappa$--symmetry of the GS formulation of the superbranes shows up from the irreducible
local worldvolume supersymmetry.}

One might have already noticed the difference in sign in the
target--space supersymmetry variations of $x^{\underline m}$
(\ref{soro-6}) and in the worldvolume supersymmetry variations
(\ref{soro-16}) and corresponding $\kappa$--variations
(\ref{soro-7}). The target--space supersymmetry and the local
worldvolume supersymmetry (or $\kappa$--symmetry) can be therefore
regarded as, respectively, `left' and `right' supertranslations of
$x^{\underline m}$.

\subsubsection{The geometrical meaning of the superembedding
condition}

To understand the superembedding condition from the geometrical
point of view let us note that the left hand side of
(\ref{soro-9}) is the Grassmann component of the pull--back onto
the superworldline of the target--space vector supervielbein
one--form
$E^{\underline a}=dX^{\underline a}-id\bar\Theta\Gamma^{\underline
a}\Theta$
\begin{equation}\label{soro-20}
E^{\underline a}|_{{\cal M}_{sw}}=(d\xi+i\eta d\eta)(\partial_\xi
X^{\underline m}-i\partial_\xi\bar\Theta\Gamma^{\underline
m}\Theta)+d\eta(DX^{\underline m}-iD\bar\Theta\Gamma^{\underline
m}\Theta),
\end{equation}
where $d\xi+i\eta d\eta$ and $d\eta$ form a basis of the
supercovariant one--forms (supervielbeins) on the superworldline.

We see that the superembedding condition (\ref{soro-9}) requires
that the pullback of the target--space {\it vector} supervielbein
$E^{\underline a}$ vanishes along the {\it Grassmann} directions of the
superworldvolume.

This is the generic requirement for the superembedding to be
appropriate to the description of the dynamics of the superbranes.

In general, if we take a supersurface ${\cal M}_{sw}$, whose
geometry is described by supervielbein one--forms $e^a(z^M)$
($a=0,1,\cdots,p$) and $e^{\alpha}(z^M)$ ($\alpha=1,\cdots n$),
and a curved target superspace $M_{TS}$, whose geometry is
described by supervielbein one--forms $E^{\underline
a}(Z^{\underline M})$ ($\underline a=0,1,\cdots,D-1$) and
$E^{\underline\alpha}(Z^{\underline M})$ ($\underline\alpha=1,\cdots
2n$)\footnote{Note that the supergeometries of ${\cal M}_{sw}$ and
$M_{TS}$ should be that of corresponding supergravities, which implies that
the torsions and curvatures of ${\cal M}_{sw}$ and $M_{TS}$ are
subject to appropriate supergravity constraints.}, and consider
the embedding of ${\cal M}_{sw}$ into $M_{TS}$, then for the
superembedding to describe a super--p--brane propagating in
$M_{TS}$ the pull--back of $E^a$ along the Grassmann directions of ${\cal M}_{sw}$
must vanish, i.e. in
\begin{equation}\label{soro-21}
E^{\underline a}|_{{\cal M}_{sw}}=e^{\underline a}E_a^{~\underline
a}+ e^{\underline\alpha}E_\alpha^{~\underline a}
\end{equation}
the Grassmann components are zero
\begin{equation}\label{soro-22}
E_\alpha^{~\underline a}\left(Z(z)\right)=0.
\end{equation}
For the most of the superbranes (with some subtleties for the
space filling and codimension one branes
\cite{soro-co1,soro-d9,soro-bppst,soro-dh}), the superembedding condition
(\ref{soro-22}), accompanied by the
$M_{TS}$ and/or ${\cal M}_{sw}$ supergravity constraints, implies
that
\begin{itemize}
\item
the geometry of the superworldvolume ${\cal M}_{sw}$ is induced by
its imbedding into $M_{TS}$, i.e. the ${\cal M}_{sw}$ supergravity
on the brane is not propagative;
\item
the dynamics of the superbrane is subject to the standard
constraints of the Green--Schwarz formulation, such as the
Virasoro constraints and their fermionic counterparts;
\item
$\kappa$--symmetry is a particular form of worldvolume
superdiffeomorphisms;
\item
the consistency of the superembedding condition results in the
same `brane scan' as that of the GS formulation.
\end{itemize}
In addition, when the number of the Grassmann directions of ${\cal
M}_{sw}$ is 16 or higher, the integrability of the superembedding
condition requires the worldvolume superfields to satisfy the
dynamical equations of motion of the superbrane
\cite{soro-bpstv}
\footnote{This is similar to the case of, say, higher
dimensional super--Yang--Mills and supergravity theories whose
superfield constraints produce the dynamical equations of
motion.}. It is in this way the covariant equations of motion of
the M5--brane were obtained for the first time \cite{soro-hsw}.

\section{Superembedding actions}
In the cases when the superembedding condition does not imply
dynamical equations but only the kinematic constraints, one can
construct (doubly supersymmetric) worldvolume superfield actions
for corresponding superbranes. Several related methods have been
proposed so far to construct the superembedding actions
\cite{soro-stv,soro-tonin,soro-other,
soro-gs,soro-hsr,soro-pst,soro-bppst,soro-dh}.

For the massless superparticles the action is simply the integral
over the superworldline of the product of the left hand side of
the superembedding condition (\ref{soro-22}) with a Lagrange
multiplier $P^\alpha_{\underline a}(z)$
\begin{equation}\label{soro-23}
S=\int d\xi d^n\eta P^\alpha_{\underline a}~E_\alpha^{~\underline
a}\left(Z(z)\right).
\end{equation}
In the case of the massive superparticles and superbranes to the
action (\ref{soro-23}) one must add a second term which governs
the dynamics of the superbrane and produces both the Nambu--Goto
(or Dirac--Born--Infeld) term and the Wess--Zumino term of the GS
formulation.

Let me explain the general structure of this second term with the
example of a space filling D3--brane in $N=2$, $D=4$ superspace.
In this case, the geometry of the worldvolume ${\cal M}_{sw}$ is
that of chiral $N=1$, $D=4$ supergravity
\cite{soro-bppst,soro-dh}.

The D3--brane couples to supergravity fields via the worldvolume
pull--back of the Wess--Zumino four--form
\begin{equation}\label{soro-WZ}
\hat C=C_4+F_2\wedge C_2+{1\over 2}F_2\wedge F_2C_0,
\end{equation}
where $F_2$ is the field strength of the BI gauge field
propagating on the D3--brane and
$C_p$ (p=0,2,4) are `Ramond-Ramond' p--form fields.

To construct the D3--brane action as an integral over the chiral
$N=1$, $D=4$ superspace ${\cal M}_{sw}$ parametrized by $z^M_L=(\xi^m_L, \eta^\alpha)$
$(\alpha=1,2)$ we take a component of the pull--back of (\ref{soro-WZ})
onto ${\cal M}_{sw}$ which has the appropriate dimensionality.
This is $\hat C_{\dot\alpha\dot\beta ab}$. We then contract $\hat
C_{\dot\alpha\dot\beta ab}$ with the antisymmetric product of the
Pauli matrices $(\sigma^{ab})^{\dot\alpha\dot\beta}$. The
D3--brane action (accompanied by the superembedding term
(\ref{soro-23})) is
\begin{equation}\label{soro-action}
S=\int d^4\xi_L d^2\eta ~{\cal E}_L
(\sigma^{ab})^{\dot\alpha\dot\beta}
\hat C_{\dot\alpha\dot\beta ab}~~+~~h.c.,
\end{equation}
where ${\cal E}_L $ is a chiral superspace integration measure
(see \cite{soro-bppst}).

Upon solving for the superembedding condition, integrating over
$\eta$ and eliminating auxiliary fields one can get from this
simply--looking superembedding action the D3--brane action of the
GS formulation. It is still a problem for future study to
demonstrate how, in the static gauge, the action
(\ref{soro-action}) is related to the super field DBI action
discussed in \cite{soro-cf,soro-pbgs,soro-bik}.

As a final remark to this section we note that when the
superembedding condition (\ref{soro-22}) produces dynamical
equations of motion and the worldvolume superfield actions cannot
be constructed there exist geometrically well--grounded recipes
\cite{soro-actions,soro-hsr} of how to construct
Green--Schwarz--type actions for corresponding superbranes.

\section{Conclusion}
We have tried to describe in simple terms basic features of the
superembedding approach, which has not only allowed one to explain
and clarify various classical and quantum properties of
superstring and superbrane theory, but has also found applications
in the construction and description of new superbrane models, and
of field theories with partially broken supersymmetry, as well as
for solving practical problems. For instance, recently it has been
used for calculating vertex operators in (M2--M5)--brane systems
\cite{soro-moore}.

 One may expect the superembedding methods to be also useful for other
 purposes, such as a unified (S--duality) description of
 fundamental and solitonic extended objects \cite{soro-ban}, and,
 in particular, for a covariant description of $N$ coincident Dp-branes and
corresponding supersymmetric non--Abelian Dirac--Born--Infeld
theories.

\begin{theacknowledgments}
I am grateful to the Organizers of the XXXVII Karpacz Winter
School for their warm hospitality, and would like to thank I.
Bars, E. Bergshoeff, M. Cederwall, J. de Azcarraga, E. Ivanov, S.
Krivonos, J. Lukierski, A. Pashnev, P. Pasti and M. Tonin for
interest to this work and discussions. This work was partially
supported by the European Commission RTN Programme
HPRN-CT-2000-00131 and by the Grant N 2.51.1/52-F5/1795-98 of the
Ukrainian Ministry of Science and Technology.
\end{theacknowledgments}

\end{document}